\documentclass[twocolumn,amsmath,amssymb,prl,showpacs]{revtex4-1}

\usepackage{graphicx,wasysym,color,SIunits}

\begin{document}
\title{Elastic instability in stratified core annular flow}
\author{Oriane Bonhomme$^{1}$, Alexander Morozov$^{2}$, Jacques Leng$^{1}$ and Annie Colin$^{1}$}
\email{annie.colin-exterieur@eu.rhodia.com}
\affiliation{$^{1}$University Bordeaux-1, Laboratory of the Future, \\178, avenue du Docteur Schweitzer, 33608  Pessac cedex -- France\\
$^{2}$SUPA, School of Physics \& Astronomy, The University of Edinburgh, JCMB, Kings Buildings, Mayfield Road, Edinburgh EH9 3JZ, United Kingdom}

\date{\today}

\begin{abstract} 

We study experimentally the interfacial instability between a layer of dilute polymer solution and water flowing in a thin capillary. The use of microfluidic devices allows us to observe and quantify in great detail the features of the flow. At low velocities, the flow takes the form of a straight jet, while at high velocities, steady or advected wavy jets are produced. We demonstrate that the transition between these flow regimes is purely elastic -- it is caused by viscoelasticity of the polymer solution only. The linear stability analysis of the flow in the short-wave approximation captures quantitatively the flow diagram. Suprisingly, unstable flows are observed for strong velocities, whereas convected flows are observed for low velocities. We demonstrate that this instability can be used to measure rheological properties of dilute polymer solutions that are difficult to assess otherwise.

\end{abstract}


\maketitle

Polymer solutions exhibit purely elastic flow instabilities even in the absence of inertia~\cite{Larson:92}. The almost ubiquitous ingredient of such an elastic instability is the curvature of streamlines: polymers that have been extended along curved streamlines are taken by fluctuations across shear rate gradient in the unperturbed state which, in turn, couples the \emph{hoop stresses} acting along the curved streamlines to the radial and axial flows and amplifies the perturbation~\cite{Shaqfeh:96,McKinley:96}. Flat interfaces between two fluids with different viscoelastic properties can also become unstable~\cite {Chen:91, Hinch:92, Wilson:97} due to the normal stress imbalance across the interface. These instabilities often occur in coextrusion where different polymers are melted in separate screw extruders and then flown simultaneously in the extrusion nozzle. Undesirable wavy interfaces are sometimes observed between the adjacent polymer layers both during the flow and in the final product~\cite{Valette:04}. Since these instabilities set severe limits to the industrial processes such as film or fiber fabrication, they have been extensively studied before \cite{Valette:04,Valette:03}. Although previous experiments and theory agree reasonably well~\cite{Han:78,Valette:04}, a comprehensive description of the flow is still lacking~\cite{Valette:04a}.

In this Letter we propose a quantitative explanation for various flow patterns observed in purely elastic interfacial instabilities. We perform a set of original experiments on co-flow of a polymer solution and water and map the full flow diagram. In contrast to previous experimental studies dealing with macroscopic flows of molten polymers, we focus here on flows of \emph{dilute polymer solutions} and water in \emph{microfluidic devices}. The advantage of using dilute polymer solutions is that their elastic properties can easily be tuned by dilution and that most of the theoretical work has been performed for models representing dilute polymeric fluids. The use of microfluidic flow geometries offers simple control and visualisation of the flow.

We observe that above some flow rate the interface between the polymer solution and water becomes wavy. Surprisingly, for relatively low velocities, the instability is convected downstream, while a stationary unstable flow is observed at high velocities. This behaviour is due to the interplay between advection by the mean flow and the growth of a perturbation and is in close agreement with the model we propose. Our work opens the route to use such an elastic instability in straight channels to promote mixing in microfluidic devices~\cite{Helton:07}. We demonstrate that it is also a new way to measure rheological properties of weakly elastic polymer solutions where very few techniques are available.

Our experiments are performed in a  microfluidic device made of nested glass capillaries: an inner capillary of square cross-section with tapered nozzle (section $\simeq 300\,\micro\meter$) nearly perfectly fits into a cylindrical capillary ($R_c = 400\,\micro\meter$) that carries the outer fluid; it offers a simple way to self-center and align the capillaries~\cite{utada:05,Guillot:07}. The length  between the nozzle and the outlet of the device is set to $L = 6\,{\centi\metre}$. The two co-flowing fluids are injected with precision syringe pumps at flow rates $Q_i$ and $Q_e$ for the internal and external rates respectively. The microfluidic chip is positioned vertically on a standard microscope (mounted accordingly) in order to prevent the effect of gravity. The observations are carried out with a fast camera (Miro Phantom). Our working fluid is a solution of Poly(VinylAlcohol) (PVA) of molar mass $M_w={196000\,\gram\per\mole}$ in water at concentrations of $3.25$, $5$, $6$ and $7.5$ $\%$ wt/wt (for which $c\gtrsim c^{\star} \approx 1 \%$ wt/wt). The measured values of the shear viscosity of the solutions are given in Table~\ref{tab:recap_time} and do not depend on the shear rate up to $10^2\,\reciprocal\second$. The viscosity of water is taken to be $\eta_i\simeq 10^{-3}~{\pascal\cdot\second}$ at {20\,\celsius}.

\begin{figure}
\includegraphics[width=0.9\linewidth]{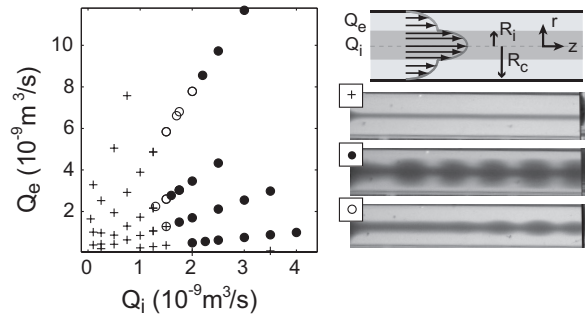}
\caption{\label{fig:PVAext} (Left) Experimental flow diagram of the $5\%$ PVA polymer solution (external fluid)
and dyed water (internal fluid) as a function of the respective flow rates. (Right) Flow patterns observed in the microfluidic chip: stable straight jets  (+), wavy jets ($\bullet$) and advected wavy jets ($\circ$).}
\end{figure}

We observe flow patterns that depend strongly on the flow rates. Fig.~\ref{fig:PVAext} shows a typical flow diagram in the ($Q_i$, $Q_e$) plane for water and the semi-dilute PVA solution as the inner and outer fluids, respectively. For low flow rates, we find straight jets (+) that extend up to the outlet of the device, while at higher flow rates the interface between the two fluids becomes unstable. At intermediate flow rates, jets are straight from the inlet up to some distance downstream where varicose undulations set in. We call them \emph{advected wavy} jets ($\circ$) and note that this distance decreases with the flow rates. At yet higher flow rates, jets are wavy through the whole set-up ($\bullet$). We note here that the advected wavy and wavy jets are, most probably, the same dynamical state, the only difference is that the latter sets in before and former after some arbitrary lengthscale $L_c$. We distinguish between the two states in order to be able to measure the relaxation time of the polymer solution, as will become apparent later. Unless mentioned otherwise, we set $L_c = 10R_c$.

In order to identify the origin of the instability, we first calculate the laminar flow profile in the system. We neglect both inertial effects and molecular diffusion processes since the Reynolds number is small ($Re\sim 0.1$) and the P\'eclet number is large ($Pe \gtrsim 10^4$) in our experiments. The two miscible fluids thus flow side by side without mixing and exhibit a constant  \emph{effective} surface tension~\cite{Korteweg:01}. The motion of water is then described by the Stokes equation while the polymer solution obeys the Oldroyd-B model~\cite{bird}:
\begin{eqnarray}
&&-\vec{\nabla} p + \eta_s \Delta \vec{v} + \vec{\nabla}\cdot{\mathbf\Sigma} = 0, \\
&&{\mathbf\Sigma} + \tau \stackrel{\nabla}{\mathbf\Sigma} = \eta_p \left( \vec{\nabla} \vec{v} + \vec{\nabla} \vec{v} ^T\right), \nonumber
\label{eq:constitutive}
\end{eqnarray}
where $\stackrel{\nabla}{\mathbf \Sigma} = \partial_t{\mathbf\Sigma} + \vec{v}\cdot\vec{\nabla}{\mathbf\Sigma}- (\vec{\nabla} \vec{v} )^T\cdot{\mathbf \Sigma} - {\mathbf \Sigma}\cdot\vec{\nabla} \vec{v}$ is the upper-convected derivative~\cite{bird}. Here, $\vec{v}$ is the velocity and $p$ is the pressure in the fluid, $\mathbf\Sigma$ is the polymer contribution to the stress tensor; $\eta_s$ is the viscosity of the solvent, and $\tau$ and $\eta_p$ are the Maxwell relaxation time of the polymer and the increase of viscosity due to the polymer chains respectively. The total shear viscosity of the polymeric solution is thus $\eta_e=\eta_p+\eta_s$. We enforce no-slip boundary conditions at the solid-liquid interface and the continuity of the velocity and of the tangential stress at the interface between the two fluids. 
In the unidirectional laminar flow of Fig.~\ref{fig:PVAext}, the pressure drop $\Delta P$ between the nozzle and the outlet of the capillary is the same in both fluids and is related by Eqs.~(\ref{eq:constitutive}) to the flow rates and the relative position of the interface between the two fluids $x=R_i/R_c$:
\begin{equation}
 \frac{\Delta P}{L}= \frac{8 \eta_e Q_e}{\pi R_c^4 (1-x^2)},
 \label{eq:dP_x}
 \end{equation}
where $x=\sqrt{\frac{\alpha-1}{\alpha-1+m}}$, $\alpha=\sqrt{1+m \frac{Q_i}{Q_e}}$, and $m = \eta_e/\eta_i$ is the viscosity ratio.

\begin{figure}
\includegraphics[width=0.9\linewidth]{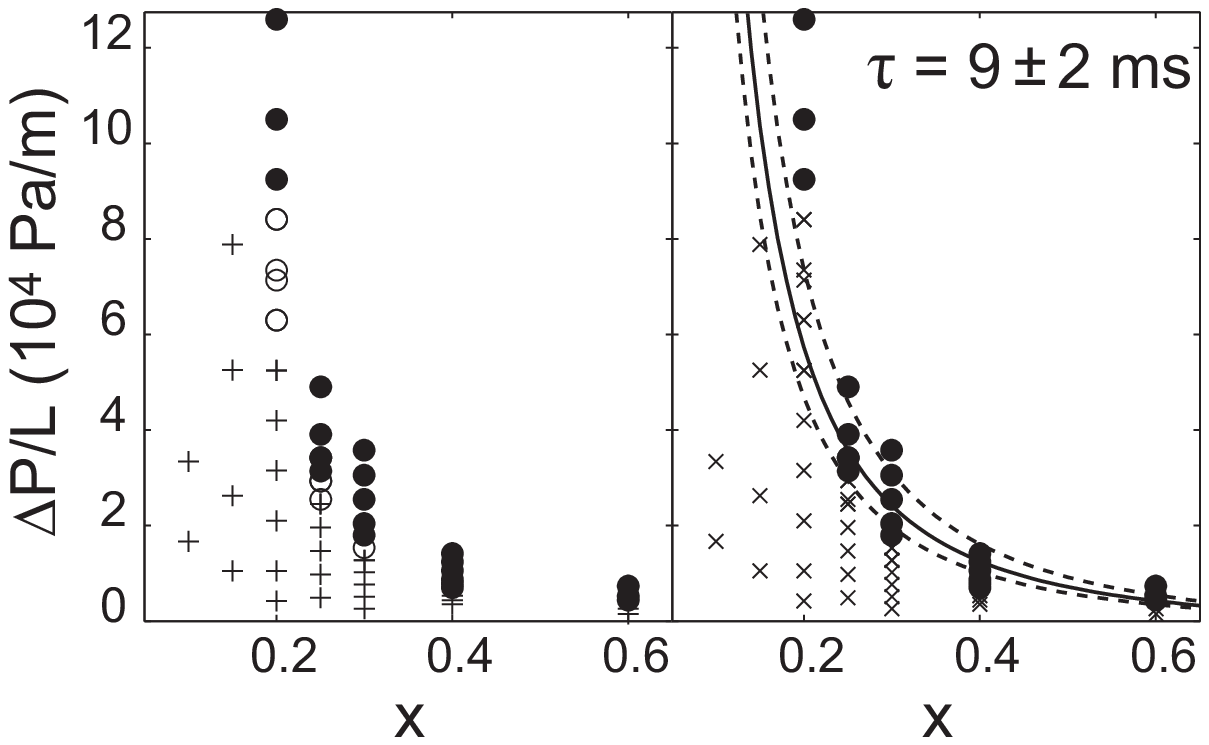}
\caption{\label{fig:xp} (Left) Flow diagram in the ($x,\Delta P/L$) plane [same data as in Fig.~\ref{fig:PVAext}: stable straight jets (+), wavy jets ($\bullet$), advected wavy jets($\circ$)]. (Right) Same data compared with the kinetic criterion~(\ref{eq:criterion}): jets with a straight part longer ($\times$) and shorter ($\bullet$) than $L_c=4.8\,{\milli\metre}$. The solid line is calculated from~(\ref{eq:criterion}) with the polymer relaxation time $\tau=9\,{\milli\second}$. The two dotted lines with $\tau=7\,{\milli\second}$ and $\tau=11\,{\milli\second}$ bracket the uncertainty.}
\end{figure}

In the left part of Fig.~\ref{fig:xp} we redraw our experimental data in the ($x,\Delta P/L$) plane and observe that stable flows occur at low while unstable flows occur at high pressure drops. This observation allows us to dismiss the Rayleigh-Plateau mechanism as a possible origin of the instability that would be triggered by the effective surface tension between the two miscible fluids~\cite{Korteweg:01,Petitjeans:96}.  Indeed, this would be contradictory to recent experimental and theoretical results~\cite{Guillot:07} that explicitly demonstrate that droplets and wavy jets (absolutely unstable states) occur at low pressure drops  while straight jets dominate at high pressure drops.

We also exclude the viscosity stratification of the flow as an origin of the instability~\cite{Hickox:71,dOlce:09}. We have repeated the same experiment with a glycerin solution with $\eta_e = 0.1~{\pascal\cdot\second}$ and \emph{we have never observed unstable inferfaces}. Since the viscosity-contrast based instability is of inertial origin, it should develop at higher Reynolds numbers. However, it does not play a role at our flow conditions.

We are thus left with a purely elastic instability driven by the constrast of the normal stresses across the interface \cite{Hinch:92}. This instability was studied analytically by Chen~\cite{Chen:91} and by Chen and Joseph~\cite{Chen:92} assuming low Reynolds and high P\'eclet numbers as in Eq.~\eqref{eq:dP_x}. They performed the linear stability analysis of the flow with respect to small axisymmetric perturbations $\propto \exp{(ikz+\omega t)}$ with $\omega$ being the complex growth rate and $k$ setting the wavelength of the perturbation. They found that long-wavelength perturbations are always stable~\cite{Chen:91}. In the opposite limit $k\rightarrow\infty$, the flow is always unstable~\cite{Chen:92} and for weak elasticity of the solution, the dispersion relation can be approximated by:
\begin{equation}\label{eq:dispersion}
\begin{split}
\omega=\frac{m(m-1)}{(1+m)^2}\left(\frac{\Delta P}{L}\frac{x R_c}{2\eta_e}\right)^2\tau
-i k \frac{\Delta P}{L}\frac{R_c^2}{4\eta_e}(1-x^2).
\end{split}
\end{equation}
An important feature of this dispersion relation is that the real part of the growth rate is independent of the wavelength. Full numerical linear stability analysis (to be published elsewhere) confirms that the dispersion curve is practically flat and positive only becoming negative for very small $k$'s. This implies that almost all wavelengths become unstable with an identical growth rate and the question of the wavelength selection cannot be answered based on the linear theory. Intriguingly, the use of a more complex rheological model would permit to raise this degeneracy~\cite{Valette:04}.
\begin{figure}
\includegraphics[width=0.9\linewidth]{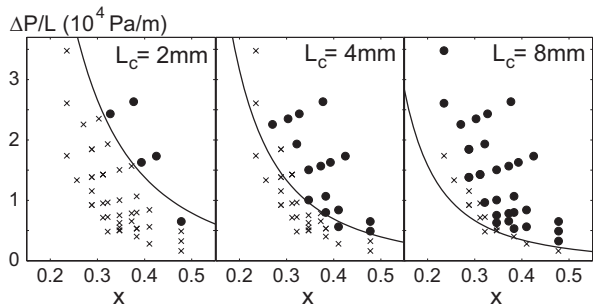}
\caption{\label{fig:variation_Lc}  State diagrams for the $3.25\,\%$wt PVA polymer solution determined for several reference distances $L_c$. A single value $\tau=8\,{\milli \second}$ is used to fit all the state boundaries.}
\end{figure}

In order to describe the convective nature of the instability we propose a \emph{simple kinetic criterion} that captures most of our experimental observations. We assume that the instability sets in very close to the nozzle and is growing on the typical timescale $\tau_r = 1/Re(\omega)$ while being advected downstream with the velocity of the interface $U$. The typical \emph{development length} of the instability is then $\tilde{L} = U\tau_r$, and the boundary between the advected wavy and wavy jets is given by $\tilde{L}=L_c$. In terms of the applied experimental parameters, this criterion reads: 
\begin{equation} 
\label{eq:criterion}
f \equiv \frac{\Delta P}{L}\frac{L_c \tau}{\eta_e} \frac{(m-1)m}{(1+m)^2} = \frac{1-x^2}{x^2}.
\end{equation}
This criterion offers a few interesting predictions. First, the dependence upon the pressure drop is quite surprising. Eq.~(\ref{eq:criterion}) implies that for a given ratio of the flow rates (which is independent of $\Delta P$), the advected wavy jets ($\tilde{L}>L_c$) occur at low pressure drops, while the wavy jets ($\tilde{L}<L_c$) should occur at high pressure drops, \emph{as observed experimentally}. Since the velocity of the interface scales linearly with the pressure drop while the destabilising forces due to the viscoelastic normal stresses scale as $\Delta P^2$ \cite{bird}, the instability moves closer to the inlet upon increase of the pressure drop. This is in contrast with the Rayleigh-Plateau instability due to the surface tension for which the opposite  order of dynamical states is observed~\cite{Guillot:07}. Moreover, the higher the pressure drop, the smaller the polymer relaxation time must be to reach the condition $\tilde{L}=L_{c}$, that turns out to be a fruitful way to measure $\tau$, as we show below. We also note that this is a purely elastic instability as it vanishes in the Newtonian limit $\tau=0$. Finally, we observe that the criterion (\ref{eq:criterion}) is indepedent of the size of the capillary that suggests that this instability will also exist in nanofluidic or macrofluidic devices provided the inertial forces are kept small.

\begin{figure}
\includegraphics[width=0.85\linewidth]{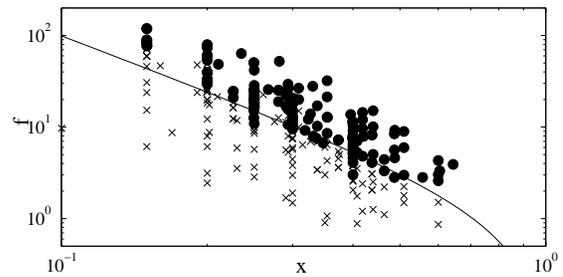}
\caption{\label{fig:exp_theory} Comparison between the experimental state diagram and the stability criterion~\eqref{eq:criterion} (solid line) for several polymer solutions (different concentrations). Solid symbols ($\bullet$) correspond to unstable flows and crosses ($\times$) to stable jets on the length $L_c$.}
\end{figure}

We now compare this kinetic criterion to our experiments. The only unknown quantity in~(\ref{eq:criterion}) is the polymer relaxation time $\tau$ and we first use it as a fitting parameter. The right panel of Fig.~\ref{fig:xp} shows that the theory based on the single-relaxation time Oldroyd-B model agrees reasonably well with the experiments. For the $5\%$ solution we have extracted $\tau=9\pm 2\,\text{ms}$. Some discrepancy however is apparent at small radii of the Newtonian core $x$. One possible source of the discrepancy is the approximate nature of the dispersion relation~(\ref{eq:dispersion}).  It is derived for short-wavelength perturbations while we use it for disturbances of intermediate wavelength. This only makes sense if the actual dispersion relation is flat for most of the $k$'s. The full numerical linear stability analysis shows that for small $x$ the dispersion relation is less flat than for large values of $x$ which possibly explains the discrepancy between theory and experiments in Fig.~\ref{fig:xp}. Another possibility is that at small $x$ we observe a decrease of the relaxation time with the local shear rate (small $x$ correspond to large pressure drops and thus high shear rates at the interface).

Next we show that the model is actually self-consistent as the measurement of $\tau$ is quite insensitive to the choice of $L_c$. In Fig.~\ref{fig:variation_Lc} we plot the stability diagram of the \emph{same system} with various reference distances $L_c$. A single value of the polymer relaxation time is able to reasonably fit most of the data. Once again, the discrepancies at small $x$ are probably due to the reasons mentioned above.

The experimental results for several polymer solutions are summarised on the master phase diagram in Fig.~\ref{fig:exp_theory}. Clearly, the simple criterion~(\ref{eq:criterion}) is remarkably successful in predicting the transition between advected wavy and wavy jets. In Table~\ref{tab:recap_time} we provide the values of $\tau$ extracted for several concentrations of the polymer. To check these values,  independent measurements of the polymer relaxation time have been performed. First, we use a version of extensional rheometry described by Amarouchene \emph{et al.}~\cite{Amarouchene:01}. When a drop of a polymer solution detaches from a capillary tube, a long-lived cylindrical neck  is formed. Initially, it thins according to a power law for all liquids, and then further, exponentially in time \emph{if the liquid is viscoelastic}. The decay time of the exponential thinning  is directly proportional to the characteristic time of the polymer $\tau_a$. 
We have also performed conventional rheological measurements of the polymer relaxation time. The data are quite noisy since the solutions are not very viscous and only weakly elastic. In Table~\ref{tab:recap_time} we report the values $\tau_w$ measured at the shear rate of $10\,s^{-1}$ which is similar to the shear rates in our coflow experiments. These values are in a reasonable agreement with the values $\tau$ extracted from the onset of wavy stationary jets. Conventional rheometry also shows that the polymer relaxation time $\tau_w$ is a decreasing function of the shear rate providing support for our explanation of the discrepancies in Figs.~\ref{fig:xp} and~\ref{fig:variation_Lc}.


\begin{table}
\caption{\label{tab:recap_time} Comparison of the polymer relaxation time obtained from the instability study ($\tau$), the drop detachment experiments ($\tau_a$) and shear rheometry done at $\dot{\gamma}=10s^{-1}$ ($\tau_w$) for several concentrations of PVA ($M_w~=196000\,{\gram \per \mole}$).} 
\begin{ruledtabular}
\begin{tabular}{l c c c c}
wt\% & $\eta_e$ ({\pascal\cdot\second}) & $\tau$ ({\milli \second})&$\tau_a$ ({\milli \second})&$\tau_w$ ({\milli \second},$\dot{\gamma}=10s^{-1}$)\\
\hline 
 7.5 & 0.65 & 120$\pm$20 & 70$\pm$ 20 & 100  \\
 6 & 0.25 & 60 $\pm$10 & 20$\pm$ 10 \\
5 & 0.1 & 9 $\pm$2 & 15$\pm$ 10 & 12\\
3.25 &0.04& 8 $\pm$2
\end{tabular}
\end{ruledtabular}
\end{table}

In this Letter, we have demonstrated that the purely elastic interfacial instability that exists between two fluids with different viscoelastic properties can not only be explained on a kinetic basis but also used to measure the polymer relaxation time in weakly elastic polymer solutions which is quite difficult to obtain otherwise. The technique also has the potential to measure shear-thinning of the solution at high shear rates. We have performed the fitting procedure described above independently for every $x$ and thus obtained the curve $\tau$ vs shear rate. The results look promising and presently we are working on the extension of the technique to more complicated constitutive relations than the Oldroyd-B model that will allow us to extract full non-linear rheology even for very dilute polymer solutions.

We would like to thank Dr. Christian Wagner and Christof Schafer, Saarland University, for their help with the rheometry. AM acknowledges support from the Royal Society of Edinburgh/BP Trust Personal Research Fellowship and EPSRC Career Acceleration Fellowship (grant reference number EP/I004262/1). LOF acknowledges support from the Aquitaine Council and from ANR Pnano project MicRheo.


%

\end{document}